\documentstyle[preprint,eqsecnum,tighten,aps]{revtex}
\def\I{{\cal I}}
\def\Schrodinger{Schr\" odinger\ }
\def\k{{\bf k}}

\def\l{{\bf l}}
\def\x{{\bf x}}
\def\y{{\bf y}}
\def\i{{\scriptscriptstyle\rm I}}
\def\f{{\scriptscriptstyle\rm F}}
\begin{document}
\draft
%\baselineskip=22pt plus 0.2pt minus 0.2pt
%\lineskip=22pt plus 0.2pt minus 0.2pt
% \font\bigbf=cmbx10 scaled\magstep3
\title{Functional evolution of free quantum fields}

\author{Charles G. Torre\thanks{torre@cc.usu.edu}}
\address{Department of Physics,
Utah State University\\
Logan, Utah 84322-4415 USA}

\author{Madhavan Varadarajan\thanks{madhavan@rri.ernet.in}}
\address{Raman Research Institute, Bangalore 560 080, India}
% \vspace*{0.25in}

\maketitle
\bigskip
\centerline{\sl November 23, 1998}
\begin{abstract}
We consider the problem of evolving a  quantum field between
any two (in general, curved) Cauchy surfaces. 
% of a flat spacetime,
% with toplogy ${\bf R}\times{\bf T}^n$, $n>1$.
 Classically, this 
 dynamical evolution is represented
by a canonical transformation on the phase space for the 
field theory.  
We show that this canonical transformation cannot, in 
general, be 
 unitarily 
implemented on the Fock space for 
free quantum fields on flat spacetimes of dimension 
greater than 2.  We do this by considering time evolution 
of a free Klein-Gordon field on a flat spacetime (with 
toroidal Cauchy surfaces) starting {}from a flat initial 
surface and ending on a generic final surface. The 
associated Bogolubov transformation is computed; it does 
not correspond to 
a unitary transformation on the Fock space. This means 
that 
functional evolution
 of the quantum 
state as originally
envisioned by Tomonaga, Schwinger, and Dirac is not a 
viable concept.
Nevertheless, we demonstrate that functional evolution 
of the quantum state 
{\it can} be satisfactorily described using the formalism 
of algebraic 
quantum field theory.
We discuss possible implications of our results for 
canonical quantum gravity.
\end{abstract}
\section{Introduction}
\label{sec:1}

In this paper we consider some aspects of 
dynamical evolution in quantum field theory. 
Specifically, we examine 
 the description of dynamics in which one evolves the state of 
 a 
quantum field {}from any initial Cauchy surface to any final 
Cauchy surface, rather than just between Cauchy surfaces of 
constant Minkowskian time. This way of formulating dynamical 
evolution 
% has been around since the 
dates back to the inception 
of relativistic quantum field theory. We begin our 
introduction to the 
main ideas via a brief historical sketch.

The idea of evolving a quantum field {}from any 
 Cauchy surface to any other  
 seems to have originated in  the mid 1940's with the work of
  Tomonaga \cite{Tomonaga}  and 
 Schwinger \cite{Schwinger} on relativistic quantum field theory. 
Tomonaga and Schwinger
wanted an invariant
generalization of the Schr\" odinger equation, which 
describes time 
evolution of the state of a quantum field relative to a 
fixed inertial reference frame.  
By allowing for all possible Cauchy 
surfaces in the description of dynamical evolution one 
easily accommodates all possible notions of time for all 
possible inertial observers. Thus a dynamical formalism 
incorporating arbitrary Cauchy surfaces 
 does allow for an
 invariant generalization of the \Schrodinger 
equation. Since, the space of Cauchy surfaces is 
infinite-dimensional, it is impossible to describe time 
evolution along arbitrary surfaces by using a single time 
parameter. 
In essence, one needs a distinct time parameter for every 
possible foliation of spacetime. 
As shown by Tomonaga and 
Schwinger, if one formulates dynamics in terms of general 
Cauchy surfaces, the resulting dynamical 
evolution equation is, formally, a functional differential 
equation, 
which is usually called the ``Tomonaga-Schwinger 
equation''. Following \cite{Kuchar2}, we use the term {\it 
functional evolution} to refer to the formulation of 
dynamical evolution in which one 
evolves quantities  along arbitrary Cauchy 
surfaces.\footnote{Synonyms for functional evolution in the 
physics
literature 
include:  
``many-fingered time'' evolution, ``hypertime'' evolution, 
and ``bubble time'' evolution.}
Thus the Tomonaga-Schwinger equation appears as 
the analog of the \Schrodinger 
equation, when describing functional evolution.
It was 
(and still is) 
tacitly assumed that the 
Tomonaga-Schwinger equation defines the infinitesimal form 
of unitary evolution of states {}from one Cauchy surface to 
another, just as the more familiar (and mathematically more 
tractable) Schr\" odinger equation describes the 
infinitesimal form of unitary evolution between two 
hyperplanes of constant Minkowskian time. One of our 
principal goals in this paper is to show that this 
assumption is untenable.

In the book {\it Lectures on Quantum Mechanics} \cite{Dirac}, 
Dirac 
considers the problem of evolving a quantum field {}from any 
 Cauchy surface to any other. He 
calls this the problem of ``quantization on curved 
surfaces''. 
He does not actually solve this problem, but rather sets up 
a constrained Hamiltonian field theory (sometimes called a 
``parametrized field theory'') that allows for evolution of 
the classical field
along any foliation of spacetime by Cauchy 
surfaces. He then considers the canonical quantization of 
this constrained Hamiltonian field theory. In this approach 
to functional evolution, the Tomonaga-Schwinger equation arises, 
formally, as the condition that constraints annihilate 
physical states. Dirac concludes
that the principal difficulty that arises is in finding a 
 factor-ordering of the operators representing energy and 
 momentum densities, 
which are to generate the dynamical evolution, so that 
the state function can be evolved consistently by the 
Tomonaga-Schwinger equation. This is essentially the 
``problem of functional evolution'', discussed by Kucha\v r 
in \cite{Kuchar2}.  Dirac indicates that this problem should 
not arise 
for fields that are ``simple'' enough, but that for field 
theories such as Born-Infeld electrodynamics the factor 
ordering problems are very severe.

As pointed out by Dirac, to obtain a 
description of the dynamics of a quantum field in 
accord with the special theory of relativity, as is 
appropriate for non-gravitational physics, 
it is not necessary to go so far as to 
allow for {\it all} Cauchy surfaces in the evolution of the 
quantum state. It is enough to secure a formulation of 
dynamics that 
allows for -- and does not distinguish between -- all notions 
of time defined by inertial observers.  
One now has only a finite number of generators to define, 
and a much weaker set of requirements to be put on 
them, namely, 
they must obey the Poincar\' e algebra.
However, when considering physical theories 
constructed in accord with the general theory of relativity, 
which is mandatory when describing gravitating systems, one 
does not (in general) have any preferred notions of time and 
one is 
obliged to consider dynamical evolution using arbitrary 
Cauchy surfaces.  Of course, it is a violation of the 
basic spirit of general relativity if one class of 
Cauchy surfaces is physically distinguished {}from another in the 
quantum theory.  Thus, in general relativistic 
physics, whether one is considering quantum fields on a 
fixed classical spacetime or considering a quantum 
gravitational field, one is naturally led to use a 
functional evolution formalism to describe the dynamics of 
the quantum field.

Motivated by issues in canonical quantum gravity, Kucha\v r 
began investigating the functional evolution 
of a free scalar field on a flat, two-dimensional 
cylindrical spacetime \cite{kuchar}. This work was continued in 
\cite{CTMV} by the 
present authors. For our purposes, the principal 
results of \cite{kuchar} and \cite{CTMV}
 are as follows. (1) Dynamical evolution 
between any pair of smooth 
Cauchy surfaces (actually, curves) is consistently and unitarily 
implemented 
on the standard Fock space representation of the quantum 
field; and as a consequence, (2) the Tomonaga-Schwinger equation 
can be rigorously defined and solved for this model.  
It was noted in \cite{CTMV} that 
generalizations of results (1) and (2) to 
flat spacetimes of higher dimension do not seem to exist. 
Our goal in this paper is to follow up on this observation. 

The extensive exploration of quantum field theory on 
a curved spacetime over the past 3 decades 
naturally has strong overlap with the 
investigation being reported here. The situation in a curved 
spacetime is, however, 
complicated by the fact that, for a generic spacetime, 
there is considerable freedom in the choice of 
Hilbert space representation of the canonical commutation 
relations (CCR) for the field. For our purposes, the most 
significant results can be found in the work of Helfer 
\cite{helfer}. 
He considers dynamical evolution of a free field 
in curved spacetime with initial and final 
surfaces being constant time surfaces in static regions (the ``in'' 
and ``out'' regions) of spacetime.  He shows 
that if one uses a Hadamard representation of the CCR
then the dynamical evolution cannot be 
unitarily implemented. Helfer shows that the putative 
generators of the functional evolution (namely, spatial 
integrals of 
certain components of 
the quantized energy-momentum tensor) necessarily have rather 
pathological 
properties. He also points out that the difficulties with 
defining the quantized generators of functional evolution already 
occur in flat spacetime.  Again, one is 
led to believe that difficulties with the functional 
evolution formalism will occur in 
spacetimes of dimension greater than 2, even  when the 
spacetime is flat.  

In what follows we will try to make this last statement 
more precise and give a ``no-go'' result for the functional 
evolution 
formalism for free fields on a flat spacetime of dimension 
greater than 2.  We begin in section \ref{sec: 2} 
by giving a formulation of the functional evolution 
formalism (for a Klein-Gordon field) using the framework of 
algebraic quantum field theory. This approach to quantum 
field theory has become more or less standard; it 
represents something of a generalization of the more 
traditional quantum field theory formalism, which is based 
upon Hilbert space, linear operators, etc. The algebraic 
approach is especially well-suited to functional evolution 
of quantum fields and, as far as we develop it here, appears 
to have  
no fundamental 
difficulties. The difficulties with functional evolution  
appear when trying to reduce the algebraic approach to the 
usual Hilbert space framework.  In particular, in 
section \ref{sec:3} we show that 
the usual Fock space for a Klein-Gordon field on a flat 
spacetime with topology ${\bf R}\times {\bf T}^{n}$ does not 
allow for unitary transformations that implement  
functional evolution.  We conclude with further 
discussion of this result and some comments on implications 
for canonical quantum gravity. Some technical details are 
collected in a couple of appendices.

\section{Algebraic Approach to Functional Evolution}
\label{sec: 2}

\subsection{Classical Preliminaries}

Here we consider the functional evolution of a free, classical
Klein-Gordon field 
$\varphi$ with mass $m$ on an $(n+1)$-dimensional, 
globally hyperbolic spacetime $(M,g)$. There is a result
due to Geroch \cite{Geroch} 
that guarantees the existence of a foliation of $M$ by 
spacelike
Cauchy 
surfaces diffeomorphic to an $n$-dimensional manifold
$\Sigma$. 
In other words, there is a diffeomorphism $\Psi\colon {\bf 
R}\times \Sigma\to M$, such that for each element of ${\bf 
R}$ the image of $\Sigma$ is a spacelike Cauchy surface.
The foliation defined by $\Psi$ is not unique, and we aim to 
describe the time evolution with respect
to an arbitary choice of 
foliation.
In  section \ref{sec:3} 
we will specialize to the 
case where $\Sigma={\bf T}^{n}$ and $g$ is flat. 

The field 
equations are
\begin{equation}
(g^{ab}\nabla_{a}\nabla_{b}-m^{2})\varphi=0.
\label{KG_eqn}
\end{equation}
We let $\Gamma$ denote the space of smooth solutions to
(\ref{KG_eqn}) with
 compactly 
supported Cauchy data. The Cauchy data are a pair
 of 
fields $(\phi,\pi)$ on $\Sigma$, where $\phi$ is a scalar field and
$\pi$ is a 
scalar density of weight one.  Given a solution $\varphi$ 
to (\ref{KG_eqn}), $\phi$ is the 
pull-back of $\varphi$ to $\Sigma$ and $\pi$ is the
pull-back 
 to $\Sigma$ 
of the normal derivative 
of $\varphi$ multiplied by the square root 
of the determinant of the induced metric on $\Sigma$.  The 
pull-back, normal, and induced metric are determined by the
embedding $T\colon \Sigma\to M$ of 
$\Sigma$ as a Cauchy hypersurface $T(\Sigma)$ in $M$. The
space of 
(smooth, compactly supported) Cauchy data will be denoted by
$\Upsilon$. 
Because the Cauchy data uniquely determine a solution, and
{\it vice versa}, the 
vector spaces $\Gamma$ and $\Upsilon$ are naturally
isomorphic once 
one has specified a Cauchy surface $T(\Sigma)$.  We denote
this 
isomorphism by $\I_{T}$. 
The map $\I_{T}\colon\Upsilon\to\Gamma$ is 
obtained by taking Cauchy data (a point in $\Upsilon$) and
evolving 
{}from the Cauchy surface $T(\Sigma)$
to get a solution of (\ref{KG_eqn}) (a point of 
$\Gamma$). The inverse map, $\I_{T}^{-1}\colon
\Gamma\to\Upsilon$, 
takes a solution to (\ref{KG_eqn}) and finds the Cauchy
data induced 
on $\Sigma$ by virtue of the embedding $T$.

Either of the spaces $\Gamma$ or $\Upsilon$ can be viewed
as the 
classical phase space for the Klein-Gordon field (see, 
e.g., \cite{Ashtekar}).  In 
particular, the
phase space in 
each case is naturally equipped with a symplectic form,
that is, a 
skew-symmetric, bilinear,  non-degenerate 2-form. 
On the 
space of solutions $\Gamma$, the symplectic form is defined
as
\begin{equation}
	\Omega(\varphi_{1},\varphi_{2})=\int_{T(\Sigma)}\sqrt{
	\gamma}\left(
	\varphi_{2}L_{n}\varphi_{1}-\varphi_{1}L_{n}\varphi_{2}
	\right),
	\label{Omega}
\end{equation}
where $L_{n}$ is the Lie derivative along the normal to the
Cauchy 
surface $T(\Sigma)$ and $\gamma$ is the determinant of the
induced 
metric on $T(\Sigma)$.  The integral is evaluated on the
Cauchy 
surface $T(\Sigma)$, but $\Omega$ is independent of the choice
of 
$T\colon\Sigma\to M$. On the space of Cauchy data
$\Upsilon$, the 
symplectic form is defined by
\begin{equation}
	\sigma((\phi_{1},\pi_{1}),(\phi_{2},\pi_{2}))
	=\int_{\Sigma}\left(\phi_{2}\pi_{1}-\phi_{1}\pi_{2}\right).
	\label{omega}
\end{equation}
It is easy to see that, for any embedding $T$, the
isomorphism 
$\I_{T}$ is a symplectic map {}from $\Upsilon$ to $\Gamma$:
\begin{equation}
	\sigma=\I_{T}^{*}\Omega.
	\label{pb}
\end{equation}

We now describe time evolution on the classical phase
space. We begin 
with evolution as represented on $\Upsilon$ since this
setting is probably most familiar.  Given initial and
final Cauchy surfaces, represented by 
embeddings $T_{\i}$ and $T_{\f}$, we view time evolution {}from 
$T_{\i}(\Sigma)$ to $T_{\f}(\Sigma)$ as a map 
$\tau_{(T_{\i},T_{\f})}\colon\Upsilon\to\Upsilon$, where
\begin{equation}
	\tau_{(T_{\i},T_{\f})}=\I_{T_{\f}}^{-1}\circ \I_{T_{\i}}.
	\label{ev1}
\end{equation}
This map  arises {}from the following 3 steps: 
(i) take initial data on $T_{\i}(\Sigma)$, (ii) evolve it to 
a solution of (\ref{KG_eqn}), and (iii) find the data that
are induced on 
$T_{\f}(\Sigma)$ by this solution. The map
$\tau_{(T_{\i},T_{\f})}$ is 
a bijection. 

Time evolution {}from $T_{\i}$ to $T_{\f}$ can also be viewed 
as a bijection ${\cal T}_{{(T_{\i},T_{\f})}}\colon\Gamma\to 
\Gamma$ on the space of solutions. We define this
map by
\begin{equation}
	{\cal T}_{(T_{\i},T_{\f})}
	=\I_{T_{\i}}\circ \I_{T_{\f}}^{-1}.
	\label{ev2}
\end{equation}
Viewed as a map on the space of solutions to the
Klein-Gordon 
equation, time evolution {}from $T_{\i}$ to $T_{\f}$ is
obtained {}from the 
following 3 steps: (i) take 
a solution to the field equations, (ii) find the data
induced on 
$T_{\f}(\Sigma)$, (iii) take that data as initial data on
$T_{\i}(\Sigma)$ 
and find the resulting solution. 

The relation between the maps $\tau$ and $\cal T$ is
through the isomorphism 
$\I$. More precisely, if we use the initial embedding
$T_{\i}$ to identify 
$\Gamma$ and $\Upsilon$, then we can use $\I_{T_{\i}}$ to
carry $\tau$ 
{}from $\Upsilon$ to $\Gamma$. We then find
\begin{equation}
	\I_{T_{\i}}\circ \tau_{(T_{\i},T_{\f})}\circ 
	\I_{T_{\i}}^{-1}
	={\cal T}_{(T_{\i},T_{\f})}.
	\label{ev12}
\end{equation}
The two descriptions of time evolution, using either $\tau$
or $\cal 
T$, are equivalent.
 Moreover, {}from the embedding independence of (\ref{Omega})
and {}from (\ref{pb}),
each is a symplectic isomorphism:
\begin{equation}
	\tau_{(T_{\i},T_{\f})}^{*}\sigma=\sigma,\quad
	{\cal T}_{(T_{\i},T_{\f})}^{*}\Omega=\Omega.
	\label{iso}
\end{equation}
This is just a field-theoretic implementation of the
familiar result 
that ``time evolution is a canonical transformation''.  

To
summarize, 
the space of solutions $\Gamma$ of (\ref{KG_eqn}) 
(or space of Cauchy data $\Upsilon$ for (\ref{KG_eqn})) 
is a symplectic vector space. Time evolution between
arbitrary Cauchy 
surfaces is a symplectic transformation on $\Gamma$
($\Upsilon$).  It 
is a matter of convenience whether we describe dynamics
using the 
space of solutions or the space of Cauchy data.  For the
most part, 
we will present our discussion using $\Gamma$, $\Omega$ and
$\cal T$.

\subsection{Quantum Field Theory}

A formulation of quantum field theory that readily allows
for 
time evolution between arbitrary Cauchy surfaces is 
provided by the algebraic approach \cite{Bratteli}. 
The main ingredients in
the 
algebraic approach are (i) 
a $C^{*}$ algebra ${\cal A}$ of basic observables and (ii) 
states, which are positive, normalized, linear functions 
$\omega\colon {\cal A}\to {\bf C}$.  The value of a state
$\omega$ on 
an observable $W\in {\cal A}$ is interpreted as the
expectation value of the 
observable represented by $W$ in the state represented by
$\omega$:
\begin{equation}
	\langle W\rangle=\omega(W).
	\label{ev}
\end{equation}

For free fields, the algebra $\cal 
A$ is taken to be the Weyl algebra, which is naturally
available on any 
symplectic vector space such as $\Gamma$ (or $\Upsilon$)
and encodes 
the information about the canonical commutation
relations.  The Weyl 
algebra is generated by elements $W(\varphi)$, labeled by
points 
$\varphi\in \Gamma$, satisfying
\begin{eqnarray}
	W(\varphi)^{*} &  & =W(-\varphi)
	\label{weyl1}  \\
	W(\varphi_{1})W(\varphi_{2}) &  & 
	=e^{-i\Omega(\varphi_{1},\varphi_{2})}W(\varphi_{1}+
	\varphi_{2}).
	\label{weyl2}
\end{eqnarray}

%For our purposes, the important feature of the Weyl algebra
%is that it 
%is essentially unique.  
Given two algebras ${\cal A}_{1}$
and ${\cal 
A}_{2}$, each with generators satisfying (\ref{weyl1}) 
and (\ref{weyl2}), there exists a unique $*$-isomorphism 
$\alpha\colon {\cal A}_{1}\to{\cal A}_{2}$ such that for
any 
$W_{1}\in {\cal A}_{1}$ and
$W_{2}\in{\cal A}_{2}$, 
we have
\begin{equation}
	\alpha\cdot W_{1}=W_{2}.
	\label{auto}
\end{equation}
This implies that {\it symplectic transformations}, that is,
linear 
transformations $S\colon \Gamma\to \Gamma$ such that
\begin{equation}
	S^{*}\Omega(\varphi_{1},\varphi_{2})
	:=\Omega(S\varphi_{1},S\varphi_{2})=\Omega(\varphi_{1}
,\varphi_{2}),
	\label{ct}
\end{equation}
are $*$-automorphisms of $\cal A$ given by
\begin{equation}
	\alpha\cdot W(\varphi)=W(S\varphi).
	\label{ctauto}
\end{equation}
 In particular, the symplectic 
transformation ${\cal T}_{(T_{\i},T_{\f})}$ representing time
evolution 
{}from $T_{\i}(\Sigma)$ to $T_{\f}(\Sigma)$ defines a
$*$-automorphism 
which we 
denote by $\alpha_{(T_{\i},T_{\f})}$.  

Time evolution in the algebraic formulation of quantum
field theory 
can be described as follows.  Assign the state $\omega$ to
the 
initial time as represented by the embedding $T_{\i}$.  Thus
the 
expectation value of the observable represented by 
$W\in{\cal A}$ on the surface
$T_{\i}(\Sigma)$ is 
given by
\begin{equation}
	\langle W\rangle_{T_{\i}}=\omega(W).
	\label{}
\end{equation}
The (inverse) automorphism
\begin{equation}
	W\longrightarrow
\alpha^{-1}_{(T_{\i},T_{\f})}\cdot W
	\label{}
\end{equation}
is the mathematical representation of time evolution of
observables in 
the Heisenberg picture.  In particular, the expectation
value of 
 $W(\varphi)$ at the final time is given by
\begin{equation}
	\langle W(\varphi)\rangle_{T_{\f}}=
	\omega(\alpha^{-1}_{(T_{\i},T_{\f})}
	\cdot W(
	\varphi))
	=\omega(W({\cal T}^{-1}_{_{(T_{\i},T_{\f})}}\varphi))
	\label{evwt}
\end{equation}

Eq. (\ref{evwt}) was built up using the Heisenberg picture. 
It is also possible to view dynamical evolution in the
\Schrodinger 
picture.  This amounts to viewing the symplectic
transformation ${\cal 
T}^{-1}_{_{(T_{\i},T_{\f})}}$ as defining a change of state rather
than a 
change in the observables.  Thus if $\omega\colon {\cal 
A}\to {\bf C}$ is the initial state, then the final state 
$\omega_{T_{\f}}$
is defined by
\begin{equation}
	\omega_{T_{\f}}=\omega\circ \alpha^{-1}_{(T_{\i},T_{\f})}.
	\label{sp}
\end{equation}
It is easy to check that $\omega_{T_{\f}}$ is indeed a state 
(positive, linear, normalized). 
Of course, in either picture the physical output of the
theory is the 
same. In particular, the expectation value of $W(\varphi)$
at the 
final time is given by (\ref{evwt}) in either picture.

The foregoing description of quantum time evolution, we
feel, is quite 
straightforward and illustrates nicely the power of the
algebraic 
approach to quantum field theory. 
Our main goal here is to consider the problem of describing
time 
evolution between arbitrary Cauchy surfaces using the more
traditional 
apparatus of Hilbert space, linear operators, and so forth.
The 
relation between the algebraic formulation and the Hilbert
space 
formulation can be made via the ``GNS construction'', which is
summarized 
as follows.  Given any state $\omega_{0}$ in the algebraic
approach, there 
exists a Hilbert space ${\cal F}$ and a cyclic
representation of the 
algebra ${\cal A}$ on $\cal F$ by bounded 
linear operators such
that the 
cyclic vector $|\psi_{0}\rangle$ is related to $\omega_{0}$ via
\begin{equation}
	\omega_{0}(W) = \langle \psi_{0}|\pi(W)|\psi_{0}\rangle,
	\label{}
\end{equation}
where $\pi(W)\colon {\cal F}\to {\cal F}$ is the operator
representative of 
$W\in{\cal A}$. The Hilbert space representation associated
to a 
given $\omega_{0}$ is unique up to unitary equivalence.  

If the underlying classical phase space is a symplectic
vector space 
of finite dimension, 
and we use the Weyl algebra to describe the basic
observables, then 
the GNS construction leads to a representation that is
unitarily equivalent to 
the standard \Schrodinger representation 
in quantum mechanics. 
% This result
% holds 
% for any choice of regular state $\omega_{0}$.
% \footnote{ A 
% regular state is one whose GNS Hilbert space yields a strongly 
% continuous irreducible unitary representation of the Weyl algebra.}
  When the
phase space is 
infinite-dimensional, unitarily inequivalent representations
can 
occur, depending upon the choice of $\omega_{0}$.  In the
case of a 
free field propagating on Minkowski spacetime, the requirements of
Poincar\' e 
invariance and positivity of energy select a unique choice
of 
$\omega_{0}$ (the vacuum state) 
{}from which the conventional Fock representation of the 
theory arises. 

Given  a Hilbert space representation of the 
Weyl algebra ${\cal A}$, 
one says that a symplectic transformation $S\colon\Gamma\to
\Gamma$, 
with corresponding algebra automorphism $\alpha\colon {\cal 
A}\to{\cal A}$ is
(unitarily) 
{\it implementable} if there is a unitary transformation
$U\colon {\cal F}\to{\cal F}$ on the 
Hilbert space $\cal F$ such that, for any $W\in{\cal A}$,
\begin{equation}
	U^{-1} \pi(W) U = \pi(\alpha\cdot W).
	\label{}
\end{equation}
Thus, implementable transformations are those which can be
represented by 
unitary operators on the chosen Hilbert space
representation of ${\cal A}$.

Thanks to the uniqueness of the Hilbert space
representation arising {}from 
 a given $\omega_{0}$ in 
the GNS construction, it follows that if $\omega_{0}$ is
invariant 
under $S$,
\begin{equation}
	\omega_{0}(\alpha\cdot W) = \omega_{0}(W),
	\label{sympinv}
\end{equation}
then $S$ is implementable on the GNS Hilbert space defined
by $\omega_{0}$.\footnote{ While (\ref{sympinv}) is sufficient for
implementability of 
$S$, it is by no means necessary.}  
In particular, if $S$ is the 
representation on $\Gamma$ of a Poincar\' e transformation,
then it is 
implementable in the standard Fock representation of the
Klein-Gordon 
field theory on Minkowski spacetime. 

On the other hand, in
contrast to 
the case where the dimension of $\Gamma$ is finite, not {\em all}
symplectic 
transformations $S$ will be implementable in field theory.
This is 
 because 
$\omega_{0}$ and its  
transform $\omega_{0}\circ\alpha$ will not always define 
unitarily equivalent Hilbert space representations. The
central issue of this 
paper is the implementability of the symplectic
transformation ${\cal 
T}_{(T_{\i},T_{\f})}$ (or, equivalently, 
${\cal T}^{-1}_{(T_{\i},T_{\f})}$ ) 
on the standard Fock representation of
a free 
field, to which we now turn.

\section{Functional evolution of a scalar field on
a flat spacetime}
\label{sec:3}

The question of unitary implementability of symplectic 
transformations on Fock representations of the
Weyl algebra 
seems to have been first answered by Shale \cite{Shale}; see 
also \cite{others}. The principal requirement for 
implementability
is that the 
mixing between creation and annihilation operators induced
by the 
symplectic transformation be described by a Hilbert-Schmidt
operator. 

Here we consider the implementability 
of functional evolution 
for a free 
Klein-Gordon field evolving on a flat spacetime.  For
technical 
simplicity we  compactify space into a torus: 
$\Sigma={\bf T}^{n}$. Standard local coordinates on $M$ 
are denoted by $X^{\alpha}$, 
$\alpha=0,1,2,\dots,n$,
where 
$X^{a}=(X^{1},X^{2},\dots,X^{n})\in (0,2\pi)$. The spacetime 
metric takes
the form
\begin{equation}
g=\eta_{\alpha\beta}dX^{\alpha}\otimes dX^{\beta},
	\label{}
\end{equation}
where $\eta_{\alpha\beta}={\rm diag}(-1,1,1,\dots,1)$.

\subsection{Classical dynamics}

Any classical solution $\varphi\in\Gamma$ to the
Klein-Gordon equation on the spacetime $({\bf R}\times{\bf 
T}^{n},\eta)$ is of the form
\begin{equation}
	\varphi=\sum_{\bf k}{1\over\sqrt{2\omega_{ k}(2\pi)^{n}}}
	\left[
	a_{\bf k}e^{i({\bf k}\cdot {\bf X}-\omega_{k}X^{0})}
	+ a^{*}_{\bf k}e^{-i({\bf k}\cdot {\bf X}-\omega_{\bf
k}X^{0})}\right],
	\label{gensol}
\end{equation}
where we use boldface to denote $n$-component vectors.
The wave 
vector $\k$ has integer components and
$\omega_{k}=\sqrt{k^{2}+m^{2}}$, 
with $k=|\k|$.  
The Fourier coefficients $a_{\k}$ and $a^{*}_{\k}$ can be
viewed as 
(complex) coordinates on the (real) phase space $\Gamma$ of
solutions to the 
Klein-Gordon equation.  
Since $\varphi$ is smooth,
the Fourier coefficients are rapidly decreasing, that is, 
$|a_{\k}|$ vanishes faster than any power of $1/k$ as
$k\to\infty$.  

The symplectic transformation ${\cal T}_{(T_{\i},T_{\f})}$
defines, and 
is defined by, a transformation of $a_{\k}$ and
$a_{\k}^{*}$.  A 
straightforward computation establishes that
\begin{equation}
	({\cal T}_{(T_{\i},T_{\f})}a)_{\k}=
	\sum_{\l} \left(\alpha_{\k\l}a_{\l} +
\beta_{\k\l}a^{*}_{\l}\right),
	\label{bog}
\end{equation}
where
\begin{equation}
	\alpha_{\k\l}=-{1\over 2(2\pi)^{n}}
	{1\over\sqrt{\omega_{k}\omega_{l}}}
	\int_{{\bf
T}^{n}}\left(\sqrt{\gamma_{\f}}n^{\alpha}_{\f}l_{\alpha}
	+\sqrt{\gamma_{\i}}n^{\alpha}_{\i}k_{\alpha}\right)
	e^{i(l_{\alpha}T_{\f}^{\alpha} -
k_{\alpha}T_{\i}^{\alpha})}\, d^{n}x,
	\label{alpha}
\end{equation}
and
\begin{equation}
	\beta_{\k\l}={1\over 2(2\pi)^{n}}
	{1\over\sqrt{\omega_{k}\omega_{l}}}
	\int_{{\bf
T}^{n}}\left(\sqrt{\gamma_{\f}}n^{\alpha}_{\f}l_{\alpha}
	-\sqrt{\gamma_{\i}}n^{\alpha}_{\i}k_{\alpha}\right)
	e^{-i(l_{\alpha}T_{\f}^{\alpha} +
k_{\alpha}T_{\i}^{\alpha})}\, d^{n}x.
	\label{beta}
\end{equation}
In (\ref{alpha}) and (\ref{beta}) we use the following
notation. The 
initial and final surfaces are described, respectively, by
the embeddings
\begin{equation}
	X^{\alpha}=T_{\i}^{\alpha}(x),\quad{\rm and}\quad
X^{\alpha}=T_{\f}^{\alpha}(x),
	\label{}
\end{equation}
where $x^{i}$ are any coordinates on ${\bf T}^{n}$.
The 
unit normal to the initial (final) Cauchy surface is 
$n_{\i}^{\alpha}$ ($n_{\f}^{\alpha}$). 
The determinant of the metric induced on the initial
(final) surface is 
$\gamma_{\i}$ ($\gamma_{\f}$).  The normals and the induced
metrics are 
functions of $x^{i}$ (and functionals of the embedding). We 
set $k_{\alpha}=(-\omega_{k},\k)$.
Our goal is to determine whether
the transformation (\ref{bog}) is implementable on the
usual Fock 
representation of the field theory, which we will now describe.

\subsection{Quantum dynamics, the Hilbert-Schmidt condition}

The principal step needed to construct an irreducible  Fock
representation of any 
linear field theory defined by a symplectic vector space 
$(\Gamma,\Omega)$ is the selection of a suitable scalar product 
$\mu\colon\Gamma\times\Gamma\to {\bf R}$ on $\Gamma$ satisfying 
\cite{Wald2,Wald}
%\begin{equation}
%	\left|\Omega(\varphi_{1},\varphi_{2})\right|
%	\leq 2\left[\mu(\varphi_{1},\varphi_{1})\right]^{1/2}
%	\left[\mu(\varphi_{2},\varphi_{2})\right]^{1/2},
%	\quad \forall\ \varphi_{1}\in \Gamma. 
%	\label{ineq}
%\end{equation}
\begin{equation}
\mu(\varphi_{1},\varphi_{1})	
= {1\over 4} \sup_{\varphi_{2}\neq 0}
                       {\left[\Omega(\varphi_{1},\varphi_{2})
                       \right]^2
                  \over 
                    \left[\mu(\varphi_{2},\varphi_{2})\right]},
	\quad \forall\ \varphi_{1}\in \Gamma. 
	\label{ineq}
\end{equation}
Given such an inner product $\mu$, 
there 
exists a (complex) Hilbert space ${\cal H}$, 
equipped with a scalar product $(\ ,\ )$, and a
real-linear 
mapping $K\colon\Gamma\to {\cal H}$ with dense range, such
that 
\begin{equation}
	(K\varphi_{1},K\varphi_{2})=\mu(\varphi_{1},\varphi_{2})
	-{
i\over 
	2}\Omega(\varphi_{1},\varphi_{2}).
	\label{HSIP}
\end{equation}
The space ${\cal H}$ is the ``one-particle Hilbert space''.
The Fock 
space arises {}from ${\cal H}$ via tensor products and direct
sums as usual.  On the Fock space the Weyl algebra is
represented 
via (densely defined, self-adjoint)
field operators built {}from creation and annihilation
operators in the 
standard way.  {}From the point of view of the algebraic
approach, the 
choice of inner product $\mu$ defines the Fock representation 
via the GNS 
construction based upon a state $\omega_{0}$, which is defined 
on the 
basic generators by
\begin{equation}
	\omega_{0}(W(\varphi)) = e^{-{1\over
2}\mu(\varphi,\varphi)},
	\label{gns}
\end{equation}
and is extended to the whole algebra by linearity and 
continuity.
In this context, (\ref{ineq}) implies 
that $\omega_{0}$ is a positive function on the Weyl 
algebra.

In terms of the expansion (\ref{gensol}), 
the scalar product $\mu$ we use to quantize the
Klein-Gordon field 
on ${\bf R}\times {\bf T}^{n}$ is given by
\begin{equation}
	\mu(\varphi_{1},\varphi_{2})={1\over
2}\sum_{\k}\left(a_{1\k}^{*}a_{2\k}
	+a_{1\k}a_{2\k}^{*}\right).
	\label{mu}
\end{equation}
This is simply the discrete momentum version of the usual
choice made in 
the standard Poincar\' e invariant quantization of the
Klein-Gordon 
field on Minkowski spacetime.  
In Appendix \ref{app:A} we indicate that the time evolution
mapping $\cal T$ is 
bounded (continuous) in this norm. 
This result is necessary for unitary implementability of the 
transformation \cite{Wald}. It also means that complications due to 
operator domain considerations do not arise.

Using the norm (\ref{mu}), 
the one particle Hilbert space $\cal H$ can be identified
with 
the complex functions
\begin{equation}
		\psi=\sum_{\bf k}{1\over\sqrt{2\omega_{ k}(2\pi)^{n}}}
	a_{\bf k}e^{i({\bf k}\cdot {\bf X}-\omega_{k}X^{0})}
	\label{wavefunction}
\end{equation}
for which
\begin{equation}
	\sum_{\k}|a_{\k}|^{2} <\infty.
	\label{}
\end{equation}
Thus $\cal H$ is the set of
  ``positive frequency'' solutions to the
Klein-Gordon 
equation with finite ``Klein-Gordon norm''.  The mapping
$K\colon 
\Gamma\to{\cal H}$ mentioned above is given by
\begin{equation}
	K\varphi=\psi,
	\label{k}
\end{equation}
with $\varphi$ defined by (\ref{gensol}) and $\psi$ defined
by 
(\ref{wavefunction}).

The bounded transformation (\ref{bog}) defines a pair of bounded
linear 
maps, $\alpha :{\cal H}\to {\cal H} $ and 
$\beta:{\cal H}\to {\bar{\cal H}}$,
where $\bar{\cal H}$ is the complex conjugate space to ${\cal H}$. 
 With $\psi$
given by (\ref{wavefunction}), 
we have
\begin{equation}
	\alpha\cdot \psi =  \sum_{\bf k,l}{1\over\sqrt{2\omega_{
k}(2\pi)^{n}}}\,
	\alpha_{\k\l}\,
	a_{\bf l}\,e^{i({\bf k}\cdot {\bf X}-\omega_{k}X^{0})}.
	\label{alphamap}
\end{equation}
and
\begin{equation}
	\beta\cdot \psi =  \sum_{\bf k,l}{1\over\sqrt{2\omega_{
k}(2\pi)^{n}}}\,
	\beta^{*}_{\k\l}\,
	a_{\bf l}\,e^{-i({\bf k}\cdot {\bf X}-\omega_{k}X^{0})}.
	\label{betamap}
\end{equation}
These transformations constitute the familiar Bogolubov 
transformation associated to any symplectic transformation on
the 
classical phase space.  The results of Refs. 
\cite{Shale,others} on unitary 
implementability state that a bounded symplectic 
transformation, such as the time evolution map $\cal T$, 
is implementable on the Fock space if and only if the
operator 
$\beta\colon {\cal H}\to {\bar{\cal H}}$ is Hilbert-Schmidt, that
is,
\begin{equation}
	{\rm
tr}(\beta^{\dagger}\beta)=\sum_{\k,\l}|\beta_{\l\k}|^{2}
<\infty.
	\label{HScondition}
\end{equation}

We now show that this condition is, in general, {\it not} 
satisfied if
the 
spacetime dimension is greater than 2.
We first give an outline of the computations, then we present
details of 
an illustrative special case. We give further
details in 
Appendix \ref{app:B}.

For simplicity, we will restrict attention to the case
where the 
initial surface is fixed and the final surface is kept 
arbitrary (that is, kept variable). 
% Of course, if the 
% functional 
% evolution is unitarily implemented, then it must be so for 
% any initial surface. Further, if evolution {}from a given initial
% surface to 
% any  final surface is unitarily implemented, then
% it follows 
% {}from the the fact that (i) unitary transformations have
% unitary 
% inverses, and (ii) compositions of unitary transformations
% are 
% unitary, that evolution between arbitrary Cauchy surfaces
% is 
% unitarily implementable.  
We 
fix the initial surface to be $X^{0}=0$ with standard
spatial 
coordinates.  Thus
\begin{equation}
	T_{\i}^{\alpha}(x)=(0,x^{a}).
	\label{Ti}
\end{equation}
We consider the validity of the Hilbert-Schmidt 
condition (\ref{HScondition}) for an
arbitrary final 
Cauchy surface.
The sum in (\ref{HScondition}) is convergent
if and only if it is absolutely convergent. Hence, to show
violation of the 
Hilbert-Schmidt condition, it suffices to show that a 
sub-sum diverges. 
Our strategy is to consider the large $k$ behavior of a 
sub-sum over $\k$
 in which ${l\over k}=O(1)$ is held fixed.
 In such a sub-sum, 
the Bogolubov coefficients $\beta_{\l\k}$ to 
leading order in $k$
 turn out to be of the form
\begin{equation}
\beta_{\l\k} =\int_{{\bf T}^n} e^{-ikG({\bf
x},\Omega_{\k})}
                                           h({\bf
x},\Omega_{\k})\,d^nx,
\label{eq:mform}
\end{equation}
where $\Omega_{\k}$ are the angular components of $\k$.
The functions $G$ and $h$ depend upon the choice of
embedding.

Eq. (\ref{eq:mform}) can be estimated by the method of
stationary phase
\cite{bleistein}. The estimate, for a large class of
embeddings 
(namely those for which the condition 
${\partial G\over \partial x^i}=0 $ is satisfied at a
finite number of
points ${\bf x}_I,$\footnote{Note 
that in general, the ${\bf x}_I,$ depend on  
$\Omega_{\k}$.}
$ I=1,2,\dots, N$, and the second
derivative matrix 
${\partial^2 G\over \partial x^i\partial x^j} $ 
has non zero determinant $D$
at ${\bf x}_I$), is 
\begin{equation}
\beta_{\l\k} = \left({2\pi\over k}\right)^{n \over 2}
    \sum_{I=1}^N f_I
                   e^{-ikG_I},
\end{equation}
where 
\begin{eqnarray}
f_I & := & |D_{I}|^{-{1\over2}}h({{\bf x}_I},\Omega_{\k}) 
e^{i\pi\,{\rm 
sign}\,D_{I}/4}
               , \\ 
G_I & := & G({{\bf x}_I},\Omega_{\k}),\quad 
D_{I}=D({{\bf x}_I},\Omega_{\k}).
\end{eqnarray}

For $n=1$, it turns out that 
$h({{\bf x}_I},\Omega_{\k})=0$ and our subsequent 
arguments for non-unitarity are not valid. But for 
generic embeddings with
$n>1$, we have
$h({{\bf x}_I},\Omega_{\k})\neq 0$, and then
\begin{equation}
\sum_{{\k}, {\l}} 
  |\beta_{\l\k}|^2 > \sum_{{\k}, \,
                                      k\rightarrow\infty}
                              {1\over k^n}
    |\sum_{I=1}^Nf_I
                    e^{-ikG_I}|^2   
            \;\;\; \sim\int^{\infty}{d^nk\over
k^n}
|\sum_{I=1}^Nf_I
              e^{-ikG_I}|^2.   
\end{equation}
Note that it is possible for $G_I =G_J$ for  some $I\neq
J$. By suitably
redefining the $f_I$, with the redefined $f_{I}$ denoted by
$\tilde 
f_{I}$, the sum 
$\sum_{I=1}^Nf_Ie^{-ikG_I}$ can be rewritten 
as a sum over a finite number ${\tilde N}\leq N$ of terms, each 
of the form 
\begin{equation}
{\tilde f}_Ie^{-ik{\tilde G}_I},\;\;\;\; I= 1,2,\dots,{\tilde
N},\;\;\; 
{\tilde G}_I\neq {\tilde G}_J\ {\rm for}\ I\neq J.              
\label{eq:generic}
\end{equation}
For generic embeddings, at least one of the ${\tilde f}_I$ is
non-zero.
We restrict attention to such embeddings. Then  
\begin{eqnarray}
\sum_{{\k}, {\l}} 
  |\beta_{\l\k}|^2 & > & 
             \int_{k\rightarrow \infty}{d^nk\over k^n}
          |\sum_{I=1}^{\tilde N}{\tilde f}_Ie^{-ik{\tilde
G}_I}|^2 \nonumber\\
               & = & \int d{\Omega_{\k}}
                     \int^{\infty} {dk\over k}( \sum_I |{\tilde
f}_I|^2 +
               \sum_{I\neq J} {\tilde f}_{I}{\tilde f}^*_{J}
                  e^{-ik ({\tilde G}_I-{\tilde G}_J)}).
\label{eq:approx}
\end{eqnarray}
The integral over $k$ of the second ($I\neq 
J$) summation converges, as
can be 
seen using integration by parts and (\ref{eq:generic}). 
The remaining integral, 
$\int {dk\over k}( \sum_{I} |{\tilde f}_I|^2)$, diverges
logarithmically with $k$
and hence $\beta$ is not Hilbert-Schmidt, that is, the 
evolution is not implementable.

In what immediately follows, we use the above type of
computation to 
demonstrate that $\beta$ is not Hilbert-Schmidt
for an illustrative special case in which the final slice
is non-trivial 
($X^{0}\neq const.$ on this slice)
but  the coordinates on it are the restrictions of the
spatial
Minkowskian coordinates, $T^{a}_{\f}(x)=x^{a}$. 
We shall refer to this as the 
``no spatial diffeomorphism'' case. The most general final
slice would 
also allow for an arbitrary set of spatial coordinates. In
Appendix 
\ref{app:B} we extend the proof to that more general case.
The discussion there has 
two parts. First we consider a case that is, in a sense,
 the exact opposite of the ``no spatial diffeomorphism''
case: the final slice is chosen to be $X^{0}=0$, so that
no real time 
 evolution has taken place, but the coordinates on the
final slice are 
 arbitrary. In other words, the final slice differs {}from
the initial 
 slice by a spatial diffeomorphism only. 
 We shall refer to this as the ``pure spatial
diffeomorphism''
case. We find that the pure spatial diffeomorphism case
also 
has the property that $\beta$ is not Hilbert-Schmidt.
Finally, we indicate how to combine 
the non-implementability results {}from the ``no spatial 
diffeomorphism'' and ``pure spatial diffeomorphism'' cases
 to the  general case of nontrivial final
slice and arbitrary spatial coordinates.
%Note that, although 
%we use Minkowskian spatial coordinates as coordinates on
%the initial slice,
% on the final slice the spatial coordinates need not be
%(the 
%restriction 
%of) the Minkowskian spatial coordinates to the slice.
%In fact we could consider the case where the final slice
%is 
%the same as the $T=0$ initial slice, but the spatial
%coordinates used are 
%not the Minkowskian spatial coordinates. 
%In such a case, the final slice and its coordinates are 
% obtained by the action of a
%spatial diffeomorphism on the initial slice. 
%

%\subsection{The `no spatial diffeomorphism' case}

In the no spatial diffeomorphism case, the initial
surface has embedding $T_{\i}^{\alpha}(x)=(0, x^a)$ and the 
final surface has embedding 
\begin{equation}
	T_{\f}^{\alpha}=(f(\x),x^{a}).
	\label{finalslice}
\end{equation}
The function $f\colon {\bf T}^{n}\to{\bf R}$ is smooth and 
satisfies
\begin{equation}
\delta^{ij}(\partial_{i}f)(\partial_{j} f)\equiv 
(\vec{\nabla}f)\cdot (\vec{\nabla}f) <1,
\label{spacelikecondition}
\end{equation}
so that the final surface is everywhere spacelike.
On the final surface we have
\begin{equation}
\sqrt{\gamma_{\f}}n_{\f}^\alpha k_\alpha= -\omega_k+
{\k}\cdot{\vec \nabla}f.
\end{equation}
Using this relation, in conjunction with an integration by
parts in (\ref{beta}),
we find (after dropping an irrelevant numerical factor)
\begin{equation}
\beta_{\l\k}
={1\over\sqrt{\omega_{k}\omega_{l}}}\,
{\omega_k\omega_l+{\l}\cdot{\k } -m^2\over \omega_k}
\int_{{\bf T}^n} 
e^{-i({\l}+{\k })\cdot {{\bf x}}+i\omega_kf}\,d^{n}x.
\label{eq:betand}
\end{equation}

We now consider the behavior of $\beta_{\l\k}$ for large 
$k$ with 
${l\over k}$ fixed and of order unity. We have
\begin{equation}
\beta_{{\l}{\k}} =\beta^{(1)}_{{\l}{\k}}+
\beta^{(2)}_{{\l}{\k}},
\label{eq:beta12}
\end{equation}
where 
\begin{equation}
\beta^{(1)}_{{\l}{\k}} :=
\int_{{\bf T}^n}
 h \, e^{-ikG}\,d^{n}x \; ,
\label{eq:beta1nd}
\end{equation}                                     
\begin{equation}
\beta^{(2)}_{{\l}{\k}} :=
\int_{{\bf T}^n}
 O({1\over k})
e^{-ikG}\,d^{n}x,
\label{eq:beta2nd}
\end{equation}                                       
with 
\begin{eqnarray}
G &:= & {{\l}+{\k }\over k}\cdot {{\bf x}} - f\; ,
\label{eq:gnd} \\ 
h &:= &\sqrt{l\over k} (1 + {{\l}\cdot{\k}\over lk})\; .
\label{eq:hnd}
\end{eqnarray}

We estimate the large $k$ behavior of 
(\ref{eq:beta1nd}) by the stationary phase method 
\cite{bleistein}. The stationary points of $G(\x)$ are
points such 
that, for given values of $\l/k$ and $\Omega_{\k}$, 
\begin{equation}
{\partial G \over \partial x^i}= 0 \;\;\; \Rightarrow \;\; 
                      {\partial f \over \partial x^i}
                      ={l_i \over k} +{k_i\over k} 
                             .
\label{eq:lnd}
\end{equation}                           

In order to allow critical points of $G$ to exist, we 
confine our attention to points in $\k$-$\l$ space such 
that $|\l+\k|^{2}<k^{2}$. 
We focus on the sub-sum of (\ref{HScondition}) obtained by 
fixing 
${\l \over k}=:\bf{L}$ and allowing  ${\k \over k}$ to vary 
in an open 
neighborhood of a fixed  unit vector $\bf K$.
To ensure that  our approximation (\ref{eq:approx}) works, we
 demand that the embedding (and the choice of  
$ \bf{L} $ and $\bf K$) be such that
the following `genericity' conditions hold.
\\
\noindent {\bf (i)} There exist only a finite 
number of points ${{\bf x}}_I, I=1,\dots,N$ on the slice where 
$\vec{\nabla} f= {\bf L}+{\bf K}$. 
\\
\noindent {\bf (ii)} the matrix 
${\partial^2 f \over \partial x^i\partial x^j}$
evaluated at ${\x}_I$ have non-zero determinant.\\
\noindent {\bf (iii)} 
$h$  given by (\ref{eq:hnd}), 
when evaluated at ${\bf l}=k{\bf L}$
and $\k =k{\bf K}$, 
is non zero.
\\
\noindent {\bf (iv)} At least one ${\tilde f}_I$ in 
(\ref{eq:generic})
be non zero.

Conditions {\bf (i)} and {\bf (ii)} imply that there 
exist a fixed, finite number of critical points of $G$ for 
each term in the sub-sum.
In particular, we note 
that at fixed $\bf L$, (\ref{eq:lnd}) defines a map 
$\lambda\colon{\bf T}^{n}\to {\bf R}^{n}$ taking $\x\in {\bf 
T}^{n}$ to ${\k\over k}\in {\bf R}^{n}$. {}From this point of view, 
{\bf (ii)} states that the 
differential of $\lambda$ is non-degenerate at the critical 
points of $G$. 
An application of the inverse mapping 
theorem to $\lambda$ implies that (a) the number of critical points 
$N$ does not change as ${\k \over k}$ varies in a 
sufficiently small
neighborhood of {\bf K}, 
and (b) the location of the  critical points varies continuously 
with ${\k \over k}$ (in a neighborhood of {\bf K}). 
The results (a) and (b) ensure that the explicit 
expression (\ref{eq:approx}) holds. Conditions {\bf (iii)} 
and {\bf (iv)} ensure that
the contribution of the critical points is non-trivial.

Any embedding which satisfies the requirements 
{\bf(i)}-{\bf (iv)}, has an
associated $\beta_1$
which is not Hilbert-Schmidt 
as can be seen by the arguments following (\ref{eq:mform}). 
The same arguments show
that 
$\beta_2$ contributes only convergent terms to 
the relevant sub-sum over $|\beta_{{\l}{\k}}|^2$. Hence,
$\beta$ is not Hilbert-Schmidt.

% We remark that the condition {\bf (i)}
% at $x_I$
% together with the spacelike 
% nature of the embedding implies $|{\bf L}+{\bf K}|<1$. 
We remark that there are two special cases where our proof of the 
failure of (\ref{HScondition}) does not apply because not all of the 
conditions {\bf (i)}--{\bf (iv)} are  satisfied. First, 
if $f$ is a constant (so that the final slice is
just a time 
translation of the initial slice, which is clearly
implementable) then 
 all of the conditions fail to be satisfied.
Second, the condition (\ref{spacelikecondition}) that the embedding 
be spacelike, in conjunction with {\bf (i)}, 
 implies that $|{\bf K}+{\bf 
L}|<1$, and it follows that ${\bf K}\cdot {\bf L}<0$. When 
$n=1$, this 
means that {\bf (iii)} is not satisfied. Indeed, it is shown in 
\cite{CTMV} that functional evolution is unitarily 
implemented when $n=1$.

% Further, if 
% the embedding is such that the only choice of ${\bf K}, {\bf 
% L}$ consistent with {\bf (i)} is 
% ${\bf L}=-{\bf K}$, then
% {\bf (iii)} is violated. In particular,
%  

% Note that for $n=1$, the argument fails because condition
% {\bf (iii)}
% is violated. In detail, . This, in
% conjunction with
% (\ref{eq:lnd}), implies that ${\l}\cdot {\k} <0$
% which in one dimension implies, {}from (\ref{eq:hnd}), that
% $h=0$. 
%However, if $n>1$ and ${\bf } \neq 0$, it is straightforward
%to show that {\bf (iii)} can always be satisfied.

When $n>1$, the conditions {\bf (i)}, {\bf (ii)}, {\bf (iv)} 
are rather weak restrictions on the final embedding.
Moreover, these conditions are probably stronger than necessary. For 
example,  if {\bf (ii)}
fails, we may expect the 
large $k$ behavior of $\beta_{\l\k}$ to be even more divergent 
\cite{bleistein}.

\section{Discussion}

We have shown that dynamical evolution of a free field on 
a flat spacetime with topology ${\bf R}\times {\bf T}^{n}$ 
will not be unitarily implemented if the initial hypersurface is 
 flat  and the final hypersurface is suitably 
generic. There is no reason to expect that the situation 
will be improved by allowing for a more general (but 
generic) initial surface. 
It is not the curvature {\it per se} of the final (and/or 
initial) 
surface which causes the problem with unitarity, but rather 
that, in general, there is no isometry of the spacetime 
that will map the 
initial surface to the final surface. Indeed, 
it is easy to see that evolution 
between any two Cauchy surfaces related by an isometry
{\it is} unitarily implemented.  
This 
situation is analogous to the Van Hove obstruction to
unitary 
implementability of the group of canonical transformations 
\cite{VH}, 
which indicates that only a subset of
the group of canonical 
transformations can be represented as unitary
transformations in quantum mechanics.  In our case, 
the canonical transformations 
defined by the isometry group of the 
spacetime are unitarily represented, but canonical 
transformations induced by more general 
mappings of the spacetime onto itself are not unitarily 
implemented.  

While we have chosen to work with a compactified model of 
space, we do not expect our result (failure of unitarity of 
functional evolution) to change if we the take the limit as the 
torus fills out ${\bf R}^{n}$.  Indeed, the failure of the 
Hilbert-Schmidt condition for $n>1$ is an ultraviolet effect 
(by computing on ${\bf R}\times {\bf T}^{n},$ we have 
tacitly ruled out infrared effects), and 
should be insensitive to the topology of the spatial manifold. 
Thus we expect that functional evolution will not be unitarily 
implemented on Minkowski spacetime.  Of course, dynamical 
evolution between any two surfaces related by a Poincar\' e 
transformation will be unitarily implemented on the 
standard free field Fock space, but this appears to be as far 
as one can go. One could attempt 
to improve this situation by considering more exotic 
representations of the CCR. If functional evolution could 
be unitarily implemented in this way, it would certainly be 
 interesting, at least mathematically. But 
physically one is forced to adopt the Poincar\' e invariant 
quantization scheme and the concomitant failure of unitarity 
of functional evolution.

The most pronounced implication of our results is that the 
Schr\" odinger picture is not available to describe 
functional 
evolution of free fields in flat spacetime using 
the traditional Fock space formulation of the quantum theory.  
The difficulties that arise in the functional evolution formalism 
are, apparently, more 
fundamental than envisioned by Dirac, who worried about 
integrability of evolution between arbitrary Cauchy surfaces. 
More precisely, Dirac was concerned with the difficulty in 
defining the Tomonaga-Schwinger equation in such a way that 
evolution {}from an initial surface to a final surface is independent 
of the choice of foliation used to interpolate between the 
initial and final surfaces.  Given that the evolution 
between initial and final surfaces is not defined by a 
unitary transformation,
 it seems impossible to make rigorous sense of 
the Tomonaga-Schwinger equation\footnote{We say this since 
the existence of a self-adjoint generator of 
time evolution, such as is needed to define the \Schrodinger 
equation, is equivalent to having a continuous 
1-parameter unitary 
group. Strictly speaking, this result does not apply to 
functional evolution since such evolution cannot be described 
as a 
1-parameter group. But one does not expect the situation 
to improve when the group assumption is 
dropped.} 
(in the \Schrodinger picture) and so issues of 
integrability of this equation appear to be academic.
We note that if
 the \Schrodinger picture of functional evolution is 
 problematic,
the Heisenberg picture is still available. It should be 
kept in mind, though, that in the Heisenberg
picture one does
not have a unitary operator to evolve the fields between 
arbitrary Cauchy surfaces.  

Unfortunately, in applications 
of the functional evolution formalism to canonical quantum gravity one 
 {\it is} working in the Schr\" odinger picture. In 
particular, the weak field and/or semi-classical limits of the 
Wheeler-DeWitt equation (or its Ashtekar variables 
counterpart) can be 
shown --- formally --- to yield a Tomonaga-Schwinger equation 
describing the 
propagation of the Schr\" odinger picture state function 
for fields on a fixed background spacetime \cite{weak}. 
This feature of the Wheeler-DeWitt equation 
has been viewed as a valuable (if formal) ``sanity check'' 
on the Dirac approach to canonical 
quantization of the gravitational field.
The 
problematic nature of the functional evolution formulation 
of dynamics in a 
Hilbert space setting indicates 
that the weak field/semi-classical limits of 
the Wheeler-DeWitt equation cannot, strictly speaking, 
be well-defined in terms 
of a Hilbert space of states. Perhaps one can even infer that
the Wheeler-DeWitt equation itself cannot be well-defined 
as an equation selecting a Hilbert space of physical 
states. After all, if one cannot sensibly describe 
functional evolution of a free field in flat spacetime 
using a Hilbert space of states, it becomes questionable 
whether 
it can be done in non-perturbative canonical 
quantum gravity. Nevertheless, our feeling is that such a 
point of view is 
overly pessimistic. Current proposals for defining a 
Hilbert space of states for quantum gravity seem very 
different {}from Fock representations of free field theory 
\cite{Rovelli}. 
Moreover, while the semi-classical and/or weak field limits may 
formally appear in terms of functional evolution of free 
fields,  there may be built-in limitations to these  
approximations {}from the underlying non-perturbative description. 
Indeed, the difficulties we have uncovered 
with functional evolution are ultraviolet problems; strong 
field/non-perturbative effects should be playing a role in 
the ultraviolet regime.

In any case, we have seen that the algebraic approach to quantum 
field theory does not appear to have any difficulties 
accommodating functional evolution, at least for free fields 
on a fixed background spacetime.  Indeed, problems 
analogous to those we encountered with functional 
evolution played a role in  motivating  
the development of the algebraic  formalism.
Thus the 
results presented here provide some
support for an algebraic approach to quantization of 
constrained Hamiltonian field theories such as arise in 
general relativity.  In such an approach the appropriate 
weak field and/or semi-classical limit would yield a 
functional evolution 
formalism such as is outlined in section {\bf II}. 

\acknowledgments

This work was supported in part by grants PHY-9600616, 
PHY-9732636 {}from the National Science Foundation.
MV thanks Rajaram Nityananda for suggesting the use of  
stationary phase methods. CGT thanks Abhay Ashtekar for 
helpful discussions related to this work.
\appendix
\section{Time Evolution is Continuous}
\label{app:A}

Here we show that the time evolution
map $\cal 
T$ on the
phase space of a Klein-Gordon field on the flat spacetime 
${\bf R}\times{\bf T}^{n}$  
 is bounded ({\it i.e.}, continuous) in the norm
$||\varphi||^{2}=\mu(\varphi,\varphi)$.  This is 
equivalent to saying that the Bogolubov transformation 
operators $\alpha$ and $\beta$ on the one-particle Hilbert 
space are bounded (continuous) maps.  This result is 
 necessary for unitary implementability 
\cite{Wald}. 
That $\cal T$ is bounded is also very convenient since 
it means that complicated operator domain issues do not arise.
  
We will establish that there is a constant $C$ such that, for
all 
$\varphi\in\Gamma$,
\begin{equation}
	||{\cal T}\varphi||\leq C ||\varphi||
	\label{bounded}
\end{equation}
To begin, it is useful to have the norm $||\cdot||$
expressed in 
terms of Cauchy 
data, that is, as a norm on the space
$\Upsilon$.  For 
simplicity, we identify $\Upsilon$ and $\Gamma$ using a
flat 
surface with standard spatial coordinates; the embedding is 
$T^{\alpha}(x)=(0,x^{a})$. 
In terms of Cauchy data $(\phi,\pi)$ on this surface we have
\begin{equation}
	||\varphi||^{2}
	=<\pi,\Lambda^{-1}\pi> + <\phi,\Lambda\phi>,
	\label{cdip}
\end{equation}
where $<\cdot,\cdot>$ is the $L^{2}$ inner-product on ${\bf 
T}^{n}$, and
\begin{equation}
	\Lambda = \sqrt{\Delta + m^{2}}.
	\label{lambda}
\end{equation}
Here $\Delta=-\delta^{ij}\partial_{i}\partial_{j}$ 
is the Laplacian defined by the induced
metric on the 
$X^{0}=0$ surface.  
A convenient form of this norm is as follows. 
Define
\begin{equation}
	u_{1}=\Lambda\phi,\quad u_{2}=\pi.
	\label{u}
\end{equation}
We now have
\begin{equation}
	||\varphi||^{2}=<u_{1},\Lambda^{-1}u_{1}> +
<u_{2},\Lambda^{-1}u_{2}>.
	\label{sobolevnorm}
\end{equation}
We write $u=(u_{1},u_{2})$.
It is easy to see that $||\cdot||$ is equivalent to a
Sobolev 
norm $||\cdot||_{-1/2}$ \cite{Taylor} on the space of fields 
$u=(u_{1},u_{2})$. 

We now proceed as follows.  We first
consider the time evolution map $\cal T$ {}{}from a flat initial 
surface to
an 
arbitrary (but non-intersecting) 
final surface. We will indicate below that this particular
time
evolution 
map is bounded in the norm (\ref{sobolevnorm}). It is easy to 
see that this
result 
implies the corresponding Bogolubov operators $\alpha$ and
$\beta$ are bounded as well.  This implies that the
Bogolubov 
operators for the inverse transformation (which are defined
in terms 
of $\alpha$, $\beta$ and their adjoints) are bounded, which
implies 
that ${\cal T}^{-1}$ is bounded. Now, the general time evolution 
map {}from $T_{1}(\Sigma)$ to $T_{2}(\Sigma)$, 
with these Cauchy surfaces arbitrary, can be obtained by
evolving
{}from 
$T_{1}(\Sigma)$ to a flat slice (using ${\cal T}^{-1}$) and then
evolving {}from 
the flat slice to $T_{2}(\Sigma)$ (using ${\cal T}$). Since both 
of these
evolution maps 
are bounded so is their composition and hence the
general 
time evolution map between any two surfaces is bounded.

The argument in the preceding paragraph reduces our task 
to showing that time
evolution is a bounded map when considering evolution 
{}from an initial Cauchy surface with 
embedding $T^{\alpha}(x)=(0,x^{a})$ to any final Cauchy 
surface with 
arbitrary embedding.  Let us introduce a foliation by 
Cauchy 
surfaces, which are labeled by $t$, that interpolates between 
the initial flat
embedding at $t=0$ and the 
final embedding at  $t=t^{\prime}$.  
We have to compare the norms of  
solutions $\varphi$ and ${\cal T}\varphi$ where the Cauchy
data at 
$t=0$ for ${\cal T}\varphi$ are the Cauchy data for
$\varphi$ at 
 $t=t^{\prime}$. By translating the Klein-Gordon equation into 
coordinates adapted to the given foliation, it is easy to
see that
to any $\varphi\in\Gamma$ there corresponds a 1-parameter 
set of  
functions $u(t)=(u_{1}(t),u_{2}(t))$, related to the Cauchy data at 
time $t$ 
via (\ref{u}), and satisfying a set of 
 strictly hyperbolic evolution equations on the chosen 
foliation (see \cite{Taylor} for the definition of strictly 
hyperbolic). The solutions of the Klein-Gordon equation,
$\varphi$ and ${\cal T}
\varphi$, respectively
correspond to solutions $u$ and ${\cal T} u$ of the 
evolution equations, where 
$({\cal T} u)(0)=u(t^{\prime})$.

{}From this line of reasoning we 
see that we want to compare the Sobolev norms $||u(0)||$
and 
$||({\cal T} u)(0)||=||u(t^{\prime})||$.
Because the evolution equations are strictly hyperbolic we
have the following estimate (valid for any Sobolev norm) 
\cite{Taylor},
$$ 
||u(t^{\prime})||\leq C||u(0)||.
$$
{}From our discussion above, this implies (\ref{bounded}). 
Thus,
time evolution {}from any initial Cauchy surface to any final
Cauchy surface is a
bounded, 
that is, continuous mapping in the norm defined by $\mu$ in 
(\ref{mu}). This also means that the norms $||\varphi||$ 
and $||\varphi||_{\cal T}\equiv ||{\cal T}\varphi||$ are 
equivalent.

\section{The pure spatial diffeomorphism case and the general case}
\label{app:B}

Here we complete the proof of failure of functional evolution 
to be implementable by 
extending our analysis from \ref{sec:3} to the 
pure spatial diffeomorphism case and the general case. 

\subsection{The pure spatial diffeomorphism case}

Here the time evolution is trivial in the sense that the 
initial and final Cauchy surfaces are identical.  The 
map $\cal T$ is obtained by considering initial and final 
embeddings that differ only by the choice of coordinates on 
the slice.  In invariant language, the two embedded 
hypersurfaces differ by a spatial diffeomorphism only.  
In this section we show that the spatial 
diffeomorphisms are not unitarily represented on the Fock 
space of a free Klein-Gordon field. 

The initial embedding is taken to be flat with standard 
coordinates: 
\begin{equation}
T_{\i}^{\alpha}(x)=(0,x^{a}).
\end{equation} 
 The final embedding 
is the same surface, but in coordinates $y^{a}=y^{a}(x)$: 
\begin{equation}
	T_{\f}(x)=(0,y^{a}(x)).
	\label{}
\end{equation}

The formula (\ref{beta}) for $\beta_{\l\k}$ takes the 
form (we drop an irrelevant constant factor)
\begin{equation}
\beta_{\l\k}
={1\over\sqrt{\omega_{k}\omega_{l}}}
\int_{{\bf T}^n} \left(-\sqrt{\gamma}\omega({\k}) 
+\omega({\l})\right)
e^{-i({\l }\cdot {{\bf x}}+ {\k}\cdot {\y})}\,d^{n}x.
%\label{eq:betapd}
\end{equation}
A coordinate change {}from $x$ to $y$ in the
integral yields
\begin{equation}
\beta_{\l\k}
=\pm{1\over\sqrt{\omega_{k}\omega_{l}}}
\int_{{\bf T}^n} \left(-\omega({\k}) 
+|\det\chi|\omega({\l})\right)
e^{-i({\l }\cdot {{\bf x}}+ {\k}\cdot {\y})}\,d^{n}y.
\label{eq:betapd}
\end{equation}
Here we view $x^{i}=x^{i}(y)$, and we have defined 
\begin{equation}
\chi^{i}_{j}:= {\partial x^i\over \partial y^j}.
\label{eq:chi}
\end{equation}
The sign of $\beta$ (which is irrelevant for the 
Hilbert-Schmidt condition) depends upon whether the spatial 
diffeomorphism is orientation preserving.

As in \ref{sec:3}, for large $k$ and ${l\over k}=O(1)$
equations (\ref{eq:beta12}), (\ref{eq:beta1nd})
and (\ref{eq:beta2nd}) hold, but $G$ and $h$ are now given
by 
\begin{eqnarray}
G &:= & {{\l}\over k}\cdot {{\bf x}}+{{\k}\over k}\cdot
{\y} \; ,
\label{eq:gpd} \\ 
h &:= &\sqrt{k\over l} (-1 + |\det\chi|\, {l\over k})\; .
\label{eq:hpd}
\end{eqnarray}
At critical points 
%$\y_{A}$, $A=1,2,\dots$, 
of $G$,
\begin{equation}
{\partial G \over \partial y^i}\Bigg|= 0 \quad \Rightarrow 
\quad
  \chi^{j}_{i}\, l_j = -k_{i}. 
\label{eq:lpd}
\end{equation}                           

As in \ref{sec:3}, we fix ${\l \over k}={\bf L}$ and vary 
${\k \over k}$
in a neighborhood
 of a fixed unit vector $\bf K$.
We demand that the 
spatial diffeomorphism
 (and the choice of ${\bf L}, {\bf K}$) 
be such that  the following 
properties hold.
\\
\noindent{\bf (i)} $\chi^{j}_{i}L_j=-K_{i}$ only at a finite
number of points ${{\bf y}}_I, I=1,\dots,N$ on the slice.\\
\noindent{\bf (ii)} The matrix 
${\partial^2 G \over \partial y^i\partial y^j}
={\partial^{2}x^{m}\over\partial y^{i}\partial y^{j}}L_{m}$ 
evaluated at ${\y}_I$ have non-zero determinant.\\
\noindent{\bf (iii)} At these points,
$h$ (given by (\ref{eq:hpd})) evaluated at $\l=k{\bf L}$,
$\k =k{\bf K}$ is nonzero.\\
\noindent{\bf (iv)} There exists at 
least one ${\tilde f}_I$ in (\ref{eq:generic})
that is non zero.\\

Any embedding which satisfies these requirements has an
associated $\beta_1$
which is not Hilbert-Schmidt 
as can be seen by the arguments following (\ref{eq:mform}). 
The role of {\bf (i)}-{\bf (iv)}
is exactly the same as in the no spatial diffeomorphism case.
Moreover, the same arguments as in the no spatial
diffeomorphism case also  show that 
$\beta_2$ contributes only convergent terms to 
the relevant sub sum over $|\beta_{{\l}{\k}}|^2$. Hence,
$\beta$ is not Hilbert-Schmidt.

We remark that if $\chi^{j}_i$ is a constant 
matrix, then not all the conditions {\bf(i)}--{\bf (iv)} are  
satisfied. In 
particular, if 
the spatial diffeomorphism 
is simply a translation (which is clearly implementable), 
then all of the conditions fail to be satisfied.
Note also that, for $n=1$, condition {\bf (iii)} is 
always violated.
To see this, use (\ref{eq:lpd}) to get (when $n=1$)
\begin{equation}
|\det\chi|(\y_{I})={k\over l},
\end{equation}
which implies that $h(y_{I})=0$.

\subsection{The general case}
Here we  combine the arguments and
 use the notation of the ``no spatial diffeomorphism'' and 
 ``pure spatial diffeomorphism'' cases.  As usual, 
 we use 
 \begin{equation}
 T^{\alpha}_{\i}(x)=(0,x^{i})\quad{\rm and}\quad
 	T_{\f}^{\alpha}(x)=(f(\x),y^{i}(\x))
 	\label{}
 \end{equation}
in (\ref{beta}). Make the
coordinate change from $x$ to $y$. Then, using the fact
that 
$\sqrt{\gamma}n^\alpha k_\alpha$ is of density weight 1
and the results of
the previous two special cases, we get
\begin{equation}
\beta_{\l\k}
={1\over\sqrt{\omega_{k}\omega_{l}}}
\int_{{\bf T}^n}
 \left(-\omega_k +{\k}\cdot{\vec \nabla}f +\chi
\omega_l\right)
e^{-i({\l }\cdot {{\bf x}}+{\k}\cdot{\y})
    +i \omega_k f({\y})}\,d^{n}x.
\label{eq:betag}
\end{equation}
Equations (\ref{eq:beta12}), (\ref{eq:beta1nd}) and
(\ref{eq:beta2nd}) hold with 
\begin{eqnarray}
G & := &  {1\over k}\left({\l}\cdot {{\bf 
x}}+{\k}\cdot{\y}\right)- f(
y)\, 
\label{gg} \\
h & := & \sqrt{k\over l}(-1 +{{\k}\over k}\cdot{\vec \nabla
}f +
                     \chi {l\over k}) \, . 
\label{eq:hg}
\end{eqnarray}
The critical points of $G$ are obtained via,
\begin{equation}
\chi_{j}^{i}{l_{i}\over k} + {k_{j}\over k} - {\partial 
f\over\partial y^{j}}=0.
\label{eq:lg}
\end{equation}                           
% and 
% \begin{equation}
% {\partial^2 g \over \partial y^i\partial y^j}= 
% \xi^{-1}_{mj} {\partial^2 x^m \over \partial y^i\partial
% y^j} 
%       ({\partial T\over \partial y^j}-{k_j\over k})-
% {\partial^2 T \over \partial y^i\partial y^j}
% \label{eq:gijg}
% \end{equation}                           

We use 
$\bf L$ and $\bf K$ as in the pure diffeomorphism and no 
diffeomorphism cases. We 
 demand that the embedding and the choice of $\bf L$ and 
$\bf K$ be such that the following properties hold.
\\
\noindent {\bf (i)} There are only  
a finite number of points 
$y_{I}$, $I=1,2,\dots, N$ where 
$\chi^i_j L_i -{\partial f\over \partial y^j}= -K_j$.
\\
\noindent {\bf (ii)} At the points $y_I$, the matrix 
${\partial^2 G \over \partial y^i\partial y^j}=
L_m{\partial^{2} x^{m}\over\partial y^{i}y^{j}}
-{\partial^{2} f\over\partial y^{i}y^{j}}$ 
has non zero determinant.\\
\noindent {\bf (iii)}  At $y_I$, 
$h$, as given by (\ref{eq:hg}), when evaluated at 
${\bf l}=k{\bf L}$ 
and ${\bf k}=k{\bf K}$ is non zero.
 \\
\noindent {\bf (iv)} At least one of the 
${\tilde f}_I$ in (\ref{eq:generic})
be non zero.

Any embedding which satisfies these requirements has an
associated $\beta$
which is not Hilbert-Schmidt. The role of {\bf (i)}--{\bf (iv)}
is the same as in the no diffeomorphism and pure diffeomorphism
cases.
Note that not all the conditions are  satisfied 
when $n=1$ or if $T_{\f}$ is a spacetime 
translation of $T_{\i}$.


\begin{thebibliography}{999}

\bibitem{Tomonaga}
S. Tomonaga,  Prog. Theor. Phys. {\bf 1}, 27 (1946).

\bibitem{Schwinger}
J. Schwinger,  Phys. Rev. {\bf 74}, 1439 (1948).

\bibitem{Kuchar2} K. V. Kucha\v r, in {\it General 
Relativity and Relativistic Astrophysics}, Proceedings of the 
Fourth Canadian Conference on General Relativity and 
Relativistic Astrophysics, edited by G. Kunstatter, D. 
Vincent, and J. Williams (World Scientific, Singapore, 1992).

\bibitem{Dirac}
P. A. M. Dirac, {\it Lectures on Quantum Mechanics},
(Belfer Graduate School of Science, Yeshiva University, New
York, 1964).

 

\bibitem{kuchar}
K. V. Kucha\v r,  Phys. Rev. D {\bf 39}, 2263 (1989).

\bibitem{CTMV}
C. G. Torre and M. Varadarajan,
 Phys. Rev. D {\bf 58}, 064007 (1998).


\bibitem{helfer}
A. Helfer, Class. Quantum Grav. {\bf 13}, L129 (1996).

\bibitem{Geroch}
R. Geroch, J. Math. Phys. {\bf 11}, 437 (1970).

\bibitem{Ashtekar} A. Ashtekar, L. Bombelli, and O. 
Reula, in {\it Mechanics, 
Analysis and Geometry : 200 Years After Lagrange}, edited by M.
Francaviglia ( North-Holland, New York 1991). 

\bibitem{Bratteli}
The literature on algebraic quantum field 
theory is quite extensive. See, for example, 
 O. Bratteli and D. Robinson, {\it 
Operator Algebras and Quantum Statistical Mechanics}, 
(Springer-Verlag, New York 1981); R. Haag, {\it Local Quantum Physics}, 
(Springer, Berlin,
1996), and references therein.

\bibitem{Shale}
D. Shale, Trans. Am. Math. Soc. {\bf 103}, 149 (1962).

\bibitem{others}
R. M. Wald, Ann. Phys. {\bf 118}, 490 (1979); 
M. Reed and B. Simon, {\it Methods of Modern Mathematical 
Physics III: Scattering Theory}, (Academic Press, New York 1979);
R. Honegger and 
A. Rieckers, J. Math. Phys. {\bf 37}, 4292 (1996).
\bibitem{Wald2}
B. Kay and R. Wald, Phys. Reps. {\bf 207}, 49 (1991).


\bibitem{Wald}
 R. M. Wald, {\it Quantum Field 
Theory in Curved Spacetime and Black Hole Thermodynamics}, 
(University of Chicago Press, Chicago 1994).


\bibitem{bleistein}
N. Bleistein and R. A. Handelsman, {\it Asymptotic 
Expansions of Integrals},
(Dover, New York, 1986).

\bibitem{VH} L. Van Hove, Acad. Roy. Belg. Bull. Cl. Sci.
{\bf 37}, 
610 (1951). Thanks to Henk Van Dam for suggesting this 
analogy.

\bibitem{weak}
For discussions of the weak field limit, see 
K. V. Kucha\v r, J. Math. Phys. {\bf 11}, 3322 (1970); 
A. Ashtekar, {\it Lectures on Non-Perturbative 
Canonical 
Gravity}, (World Scientific, Singapore 1991). For a 
review of the semi-classical approximation in quantum 
gravity, see E. 
Alvarez, Rev. Mod. Phys. {\bf 61}, 561 (1989), ref. 
\cite{Kuchar2}, and 
references therein.

\bibitem{Rovelli}
See, for example, C. Rovelli, {\it Living Reviews in 
Relativity}, 
{\bf 1}, 1998-1 (1998).
{\tt 
http://www.livingreviews.org/Articles/Volume1/1998-1rovelli/}


\bibitem{Taylor} M. Taylor, {\it 
Pseudo-differential Operators and Nonlinear PDE},
(Birkhauser, Boston, 1991).

\end{thebibliography}
\end{document}